\documentclass[ aip,
 jmp,
 amsmath,amssymb,
  reprint,  ]{revtex4-1}

\usepackage{graphicx}
\usepackage{subcaption}
\usepackage{hyperref}
\usepackage[inline]{enumitem}
\usepackage[dvipsnames]{xcolor}
\definecolor{backgray}{gray}{0.9}
\usepackage{listings}
\newcommand{\EPICS}[0]{EPICS}\newcommand{\main}[0]{\texttt{main()}}
\newcommand{\run}[0]{\texttt{run()}}
\newcommand{\node}[1]{\textsf{#1}}
 
\begin{document}

\title{ Distributed State Machine Supervision for Long-baseline
  Gravitational-wave Detectors }

\author{
  Jameson Graef Rollins
}
\email{jameson.rollins@ligo.org}
\affiliation{LIGO Laboratory, California Institute of Technology, Pasadena, CA 91125, USA}

\date{\today}

\begin{abstract}
  The Laser Interferometer Gravitational-wave Observatory (LIGO)
  consists of two identical yet independent, widely-separated,
  long-baseline gravitational-wave detectors.  Each Advanced LIGO
  detector consists of complex optical-mechanical systems isolated
  from the ground by multiple layers of active seismic isolation, all
  controlled by hundreds of fast, digital, feedback control systems.
  This article describes a novel state machine-based automation
  platform developed to handle the automation and supervisory control
  challenges of these detectors.  The platform, called
  \textit{Guardian}, consists of distributed, independent, state
  machine automaton nodes organized hierarchically for full detector
  control.  User code is written in standard Python and the platform
  is designed to facilitate the fast-paced development process
  associated with commissioning the complicated Advanced LIGO
  instruments.  While developed specifically for the Advanced LIGO
  detectors, Guardian is a generic state machine automation platform
  that is useful for experimental control at all levels, from simple
  table-top setups to large-scale multi-million dollar facilities.
\end{abstract}
 
\maketitle

\section{Introduction}

The Laser Interferometer Gravitational-wave Observatory (LIGO) has
just completed its first observing run with its new second-generation
instruments.  During this run, Advanced LIGO made the first ever
direct observation of gravitational waves from a binary black hole
merger~\cite{GW150914,GW151226,arxiv:1606.04856}.  This detection
heralds a new era of gravitational wave astronomy, where gravitational
wave detectors like LIGO will operate as true astronomical
observatories, continuously listening to the cosmos for
gravitational-wave events.

Achieving the scientific goals of LIGO requires robust detector
operations.  LIGO consists of two identical yet independent 4-km
baseline Michelson-type interferometric detectors, located in
Livingston, LA, and Hanford, WA, USA~\cite{Aasi_2015,Abbott_2016_1}.
The detectors directly measure the strain of passing gravitational
waves as they modulate the length of the two Michelson arms.  The
detectors are complex opto-mechanical systems consisting of multiple
subsystems whose actions and configurations need to be coordinated to
acquire and maintain the operating configuration needed to detect
gravitational waves.  Getting the detectors to their operating point
quickly and robustly is critical for maximizing observation time and
scientific output.

While many large physics experiments rely on some form of real-time
control at the machine interface level to handle fast event sequencing
(e.g. in particle accelerators or extreme light sources), LIGO is
somewhat unique in the prominent role that feedback plays in overall
detector control.  In general, though, regardless of the low-level
controls architecture, all large physics experiments require some kind
of supervisory control layer to coordinate actions between subsystems
and to achieve global operating configurations from ``cold boot''
conditions.

In Initial LIGO~\cite{doi:10.1088/0034-4885/72/7/076901}, detector
feedback control systems were simple enough to be supervised by a
small set of shell scripts and a single daemon program that monitored
detector state and executed the appropriate scripts as needed.
Advanced LIGO, however, is significantly more complex than Initial
LIGO, employing roughly 100 times the number of feedback control
loops.  The Initial LIGO supervision system was therefore inadequate
to meet the needs of Advanced LIGO.

The Virgo project~\cite{virgo}, which operates an interferometric
gravitational wave detector similar to those used by LIGO, developed
its own unique supervisory system, called \textit{Alp}, to handle
automation of their first generation
detectors~\cite{virgo,virgo-icalepcs}.  Alp, while successful in
automating the initial Virgo detectors, is specific to the inner
workings of the Virgo data acquisition system and relies on a custom
network communication layer.  It is unclear how much this system will
be used for the Advanced Virgo project~\cite{avirgo}, which is now in
its commissioning phase.

The standard for industrial automation and slow control are
programmable logic controllers (PLCs), typically embodied by the IEC
61131 standard~\cite{iec61131-3}.  Because of their ubiquity in
industry they are frequently used in large physics experiments as
well~\cite{Bauer_2012, Lagin}.  While they can be used for
higher-level supervision tasks, PLCs are usually fully integrated
systems that are less well suited for large distributed applications
that frequently require operation under some level of partitioning.
They are therefore generally reserved for low-level slow controls and
machine protection.

\begin{figure*}[!htb]
  \includegraphics[width=\textwidth]{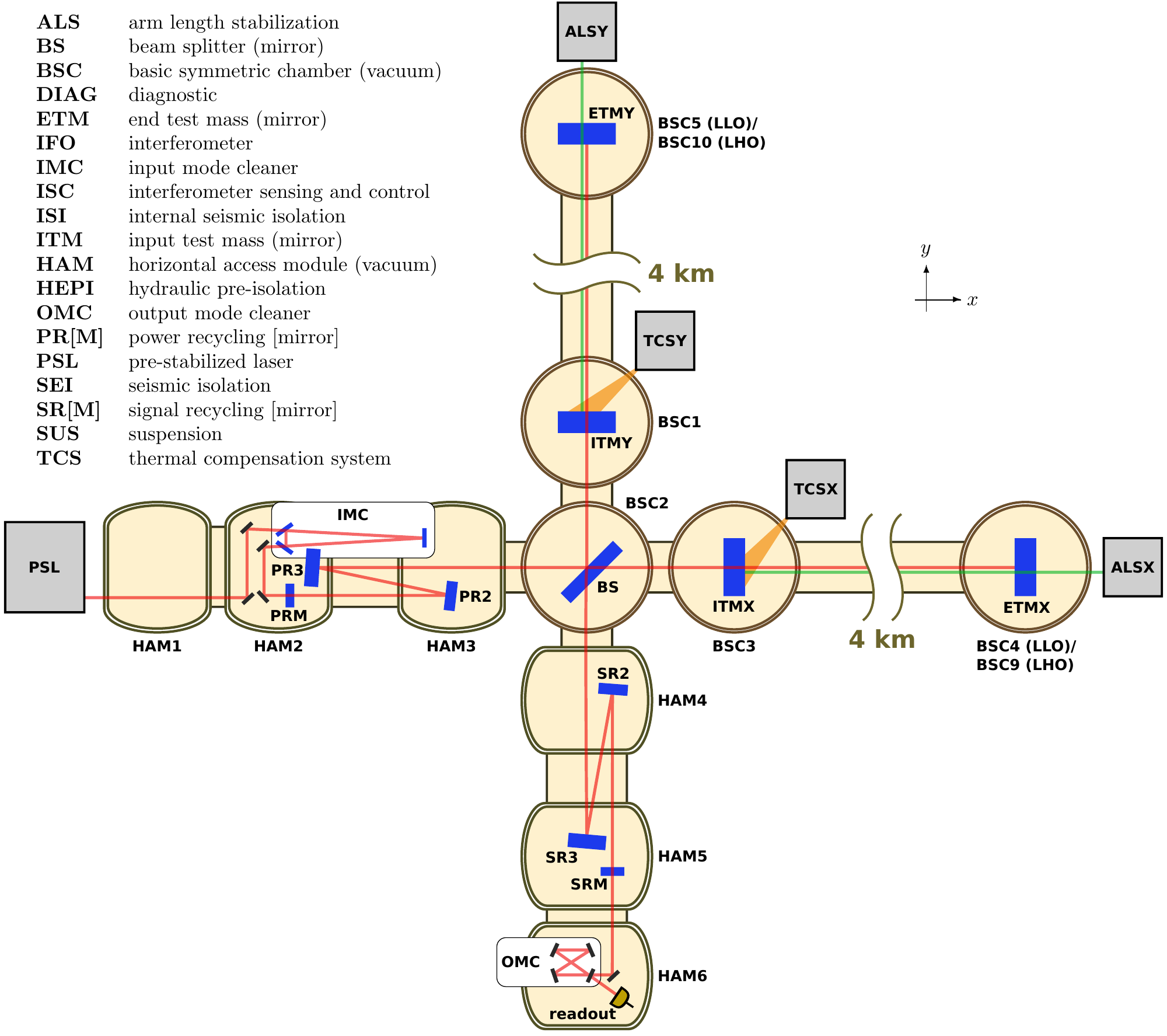}
  \caption{Overview of the Advanced LIGO vacuum envelope (tan) and
    optical configuration, and glossary.  The circular vacuum chambers
    are called BSC chambers, and the rectangular ones are called HAM
    chambers.  All vacuum chambers contain active seismic isolation
    (SEI) platforms that support the optics and their suspension (SUS)
    structures (represented by blue rectangles).  The red lines
    represent the path of the primary laser beam (emanating from the
    PSL), while the green lines represent the laser beams for the arm
    length stabilization (ALS) system.  Illuminating the input test
    masses (ITMX and ITMY) is the thermal compensation system (TCS).}
  \label{fig:ifo}
\end{figure*}

A class of supervisory control and data acquisition (SCADA) systems
has been developed specifically to meet the needs of large physics
experiments.  The primary examples of these systems are the
Experimental Physics and Industrial Control System
(\EPICS)~\cite{epics,epics-icalepcs} and
TANGO~\cite{tango,arxiv:cs/0111028}.  These systems are designed for
distributed control of large numbers of independent devices and
typically include network message passing infrastructures as well as
sequential logic programming tools for device-level automation.  LIGO
relies on EPICS as the primary communication layer for supervisory
control.  However, while these SCADA systems provide suitable mediums
for supervisory control, they don't provide much in the way of
structure or functionality for the development and management of
higher level supervision tasks.

A common model used to represent automation systems is the
\textit{finite state machine}.  Finite state machines are naturally
represented by graphs, where states corresponding to particular
configurations of, or commands on, a system are represented by nodes
in the graph, connected together by directed edges defining allowable
transitions between states.  Finite state machine representations are
quite powerful and intuitive and are well suited for many automation
tasks.  Finite state machines are found in various experiments at the
Large Hadron Collider (LHC)~\cite{doi:10.1088/1742-6596/219/2/022031,
  doi:10.1109/23.710969, alice-icalepcs, Bauer_2012,
  Misiowiec:1392942}.

In order to meet the unique supervisory requirements of the Advanced
LIGO detectors, LIGO has developed a novel state-machine based
automation platform, known as \textit{Guardian}.  Guardian consists of
a distributed hierarchy of independent automaton processes.  Each
automaton process is a state machine execution engine handling control
of a particular sub-domain of the full system.  A hierarchy of nodes
control the full instrument.  Guardian's highly distributed
architecture allows large systems to be easily partitioned and
re-unified as needed.  It is also scalable, able to handle systems
from a single automaton up to the hundreds required for large
facilities like LIGO.  Guardian was designed to be flexible and
accessible, which is important for facilitating the unique
commissioning process of long-baseline interferometers where the full
automation procedure is not known \textit{a priori} and automation
logic changes need to be made quickly and frequently.

This article describes the Advanced LIGO requirements that led to the
development of this new platform, the technical design of Guardian
itself, and how Guardian was deployed to fully control the Advanced
LIGO detectors.

\section{Advanced LIGO system description and automation requirements}

\begin{figure*}
  \includegraphics[width=\textwidth]{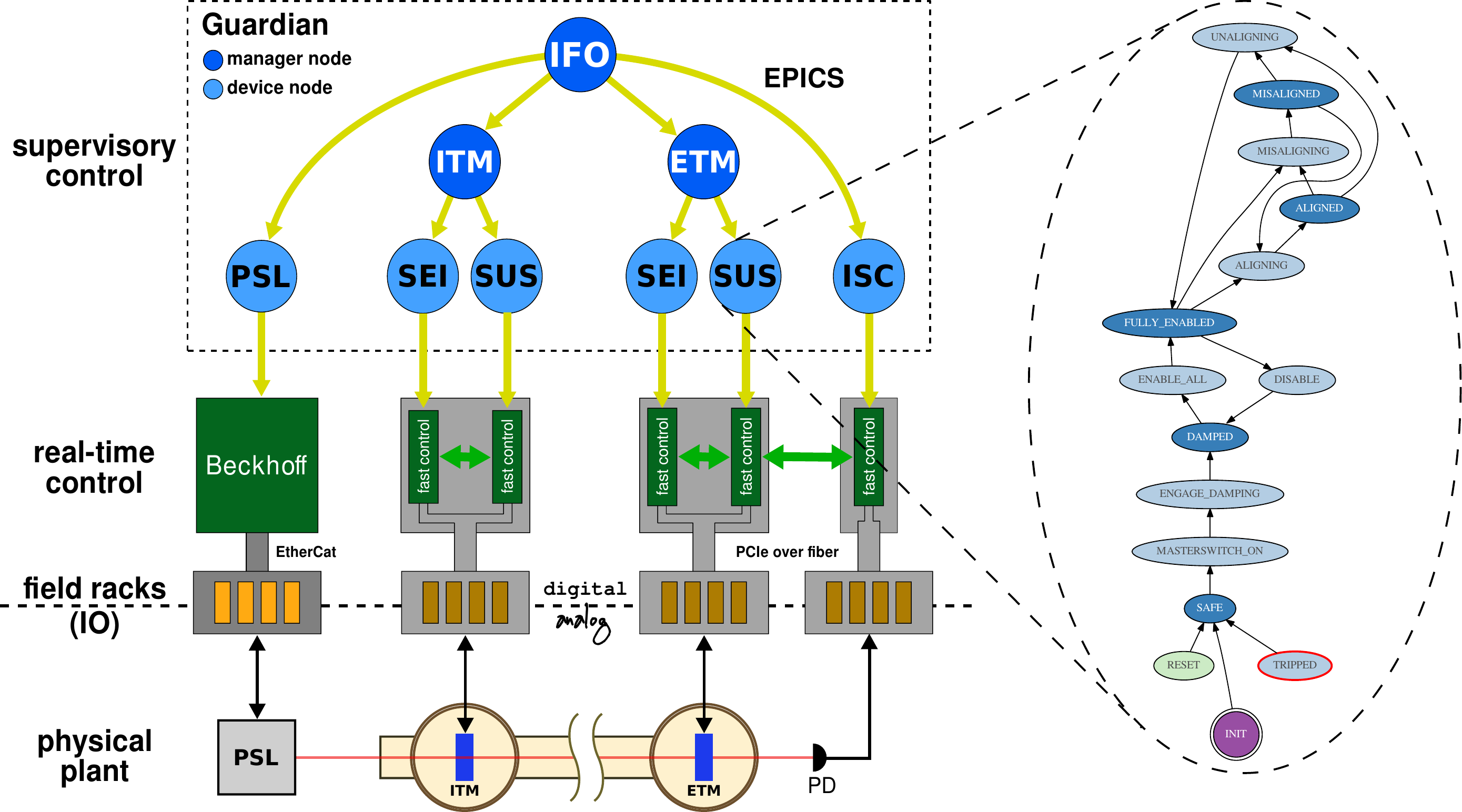}
  \caption{Overview of the Advanced LIGO digital control and
    supervision architecture.  The LIGO interferometer itself is
    represented by the ``physical plant'' layer at the bottom.  The
    input/output and ``real-time'' digital controls (also referred to
    as ``front end'' systems) are represented by the gray boxes in the
    middle layer.  Supervisory control is handled by Guardian, which
    is shown in the dashed box at the top.  The blue circles represent
    individual Guardian nodes (labeled with hypothetical subsystem
    names).  All communication between Guardian nodes and between
    Guardian and the front-end systems is handled via \EPICS.  At the
    right is an example state graph for one of the nodes.}
  \label{fig:overview}
\end{figure*}

At the core of the Advanced LIGO detectors is a dual-recycled
Michelson interferometer with Fabry-P\'{e}rot arm
cavities~\cite{Aasi_2015,Abbott_2016_1}.  To isolate the
interferometer from ground motion, all main optical components are
suspended by pendulum systems hung from active seismic isolation
platforms and can be actuated on in angle and length.  The primary
laser light source provides up to 180 Watts of input light power to
reduce the effects of quantum shot noise at the detector.  A thermal
compensation system uses various thermal actuators to counteract the
effects of laser heating in the core optics.  Figure~\ref{fig:ifo}
shows a schematic overview of the detector, as well as a glossary for
abbreviations and acronyms used in this article.

Figure~\ref{fig:overview} shows a cartoon overview of the Advanced
LIGO control and supervision architecture.  Signals from sensors in
the interferometer (labeled ``physical plant'' in the diagram) are
digitized and fed into a custom real-time control
system~\cite{dcc:T0900612}.  From the digitized error signals the
controls system calculates control signals that are fed to actuators
that affect the interferometer and its various subsystems.  A Beckhoff
industrial PLC-based control system~\cite{beckhoff} handles low-level
slow control of some discrete components of the system.

The dashed box in the top of the diagram in Figure~\ref{fig:overview}
encloses the supervisory control layer.  The real-time control system
exposes signal readbacks and parameters of the fast control as
channels in the \EPICS\ network message passing
infrastructure~\cite{epics,epics-icalepcs}.  These channels are made
available to automation and supervisory control processes
(e.g. Guardian, shown as blue circles), as well as to human operator
interfaces.  Compatibility with \EPICS\ was a fundamental requirement
for the Advanced LIGO automation system.

\subsection{Lock acquisition and global control}

The nominal operating point of the instrument---where all interference
conditions in the interferometer are such that it is maximally
sensitive to differential length changes of the Michelson arms---is
susceptible to external disturbance and must be maintained via fast
feedback control loops that ``lock'' all global length and angular
degrees of freedom to their desired set points~\cite{Staley_2014}.
The subsystem that handles all interferometer global degrees of
freedom is called interferometer sensing and control (ISC).

To achieve lock, the overall global control system and all detector
subsystems must progress through sequences of states in a well
orchestrated manner.  Initially independent components must be
controlled to progressively tighter degree by increasingly
interdependent control loops.  This process is known as \textit{lock
  acquisition}.  Developing the lock acquisition procedure is one of
the more difficult challenges in the commissioning of long-baseline
gravitational wave detectors and is the primary automation task during
operation.

Multiple detector subsystems are either directly involved in overall
continuous global control or are in some way involved in the lock
acquisition process: power levels are set depending on the current
lock state (PSL~\cite{Kwee_2012}, TCS~\cite{arxiv:1608.02934}); optics
are continuously actuated on for stability and global control (SUS,
SEI/HEPI/ISI); various sub-cavities are locked at different stages
(ISC, IMC~\cite{Mueller_2016},
ALS~\cite{Mullavey_2011,doi:10.1364/JOSAA.29.002092},
OMC~\cite{Fricke_2012}).

Eventually something will cause the global interferometer control
loops to lose control authority (i.e. ``lose lock'').  This is usually
due to external disturbances that cause controller outputs to run into
dynamic range limits.  The automation system must be able to identify
that a lock loss has occurred, reset all controllers and subsystems
appropriately, and then reacquire lock as fast as possible so as to
minimize downtime.

\subsection{Suspensions and Seismic isolation}
\label{sec:sei}

The Advanced LIGO test masses are isolated from the ground via seven
stages of active and passive seismic isolation, distributed among
three discrete subsystems: test mass/optic suspensions (SUS), internal
seismic isolation (ISI) platforms, and hydraulic external
pre-isolation (HEPI) systems.  The terms ``external'' and ``internal''
in this context are relative to the vacuum envelope.  The HEPI and ISI
systems together constitute the overall seismic isolation (SEI)
subsystem.

All test masses and auxiliary optics are suspended from multi-stage
pendulum suspensions.  The core test masses are hung from four-stage
suspensions~\cite{Robertson_2002} while the various auxiliary optics
are hung from three-, two-, and single-stage suspensions.  The
suspensions incorporate various sensors and actuators on multiple
different stages that are used for local motion sensing.  Magnetic and
electrostatic actuators are used for pushing on the stages to control
the pitch, yaw, and longitudinal degrees of freedom.  The suspension
systems are the primary actuators for global control of the full
interferometer.  They also use feedback from the local sensors to damp
various mechanical modes.

The ISI systems provide either one or two stages of active isolation
for six degrees of freedom~\cite{Matichard_2015}.  They are located
inside the vacuum enclosure and directly support the test mass and
auxiliary optic suspension structures.  There are five two-stage ISIs
for the five core test masses (BS, two ITMs, and two ETMs), and five
single-stage ISIs for five of the auxiliary optic chambers (HAMs 2-6).

The HEPI systems are located outside the vacuum enclosure between the
primary support pillars attached to the ground and the cross beams
that support the ISI platforms in vacuum.  They provide gross DC
alignment and low-frequency isolation for six degrees of freedom using
hydraulic actuation~\cite{Matichard_2015}.  There is one system for
every vacuum chamber in the system, for a total of 11 units in the
full system.

The HEPI and ISI systems have similar control system architectures for
each individual stage.  They both utilize damping loops to damp rigid
body modes and structural resonances.  The ISI systems additionally
have isolation loops for inertial isolation of the payload from input
ground motion.  These control loops must be engaged in a specific
order to maintain stability: damping loops must be engaged first for
all stages, moving from outer stages to inner, followed by the
isolation loops which must be engaged in the same order.

All SUS and SEI systems include software watchdogs that automatically
shut off all actuator outputs if control signals surpass certain
thresholds.  Operator intervention is required to reset the watchdogs,
after which all control loops must be reengaged.

\subsection{User interface}

While the lock acquisition sequence and controls necessary to realize
low-noise operation are understood in principle beforehand, the
ultimate implementations used during operation are discovered over the
course of an intensive multi-year commissioning process.  Any
supervision system must support the fast turn-around pace of
commissioning.  The system should have a clean, standardized interface
and be capable of incorporating code changes quickly and robustly so
that new ideas can be tested at a fast pace.

\section{Guardian state machine supervision}
\label{sec:guardian}

Many options were considered when looking for a supervisory solution
for Advanced LIGO.  The primary requirement was that the system work
with the existing \EPICS\ framework in the real-time control system.
This restriction narrowed down the options considerably.  The solution
also needed to be scalable to handle a large number of independent
components and flexible enough to allow partitioning of the system so
that sub-components could be commissioned independently.  This pointed
to a distributed architecture that would allow separate systems to
function independently, yet unify as a whole for full detector
control.  Finally, the system needed to support fast turnaround for
code changes---lengthy recompilation should not be necessary to
incorporate new logic.  Ultimately it was determined that no existing
system fit these requirements, leading to the development of an
entirely new system: \textit{Guardian}.

The basic concept of Guardian is that of a distributed hierarchy of
state machine automata.  Each automaton handles control of a specific
sub-domain of the full system to be controlled, and a hierarchy of
automata can be used to control larger systems.

The core of Guardian---the automaton---is the \texttt{guardian}
program, a stand-alone state machine execution engine with an \EPICS\
control interface.  When launched, the program loads user code that
defines a state graph for the system being controlled (see
section~\ref{sec:graph}).  The program knows how to traverse the
user-defined state graph to move from whatever state the system is in
to a desired final state.

In a typical large-scale deployment, such as that of Advanced LIGO, a
hierarchy of \texttt{guardian} processes---each referred to
individually as a \textit{node}---control the entire instrument
(henceforth referred to as the \textit{plant}).  Higher-level
\textit{manager} nodes control lower-level \textit{subordinate} nodes,
down to \textit{device} nodes that talk directly to the front-end
systems.  The blue circles in the top box in Figure~\ref{fig:overview}
each represent a node in a full hierarchy.  Figure~\ref{fig:hier}
shows two actual sub-hierarchies used in the full Advanced LIGO
Guardian implementation.

All Guardian user code is written in standard Python~\cite{python} (as
is the core \texttt{guardian} program itself).  The full suite of
standard Python libraries are available to the user code, as are
special libraries designed specifically for interacting with the LIGO
control system.

\subsection{State graphs and system dynamics}
\label{sec:graph}

\begin{figure}
  \begin{subfigure}[b]{0.45\columnwidth}
    \includegraphics[height=30em]{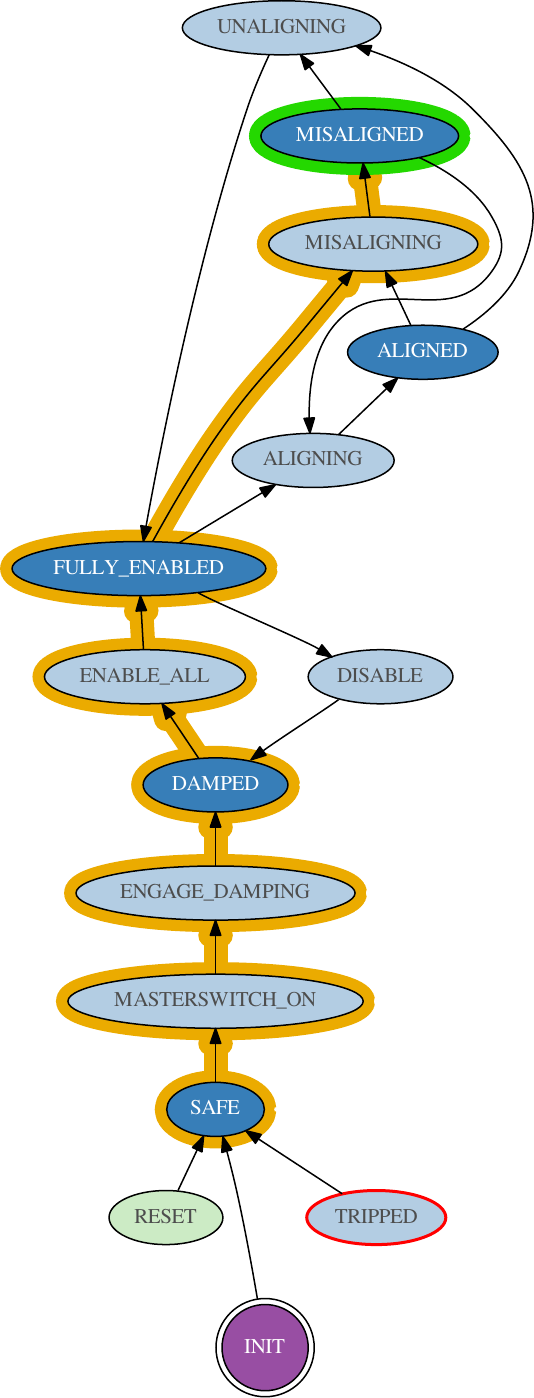}
    \caption{path}
    \label{fig:graph:path}
  \end{subfigure}
  \begin{subfigure}[b]{0.49\columnwidth}
    \includegraphics[height=30em]{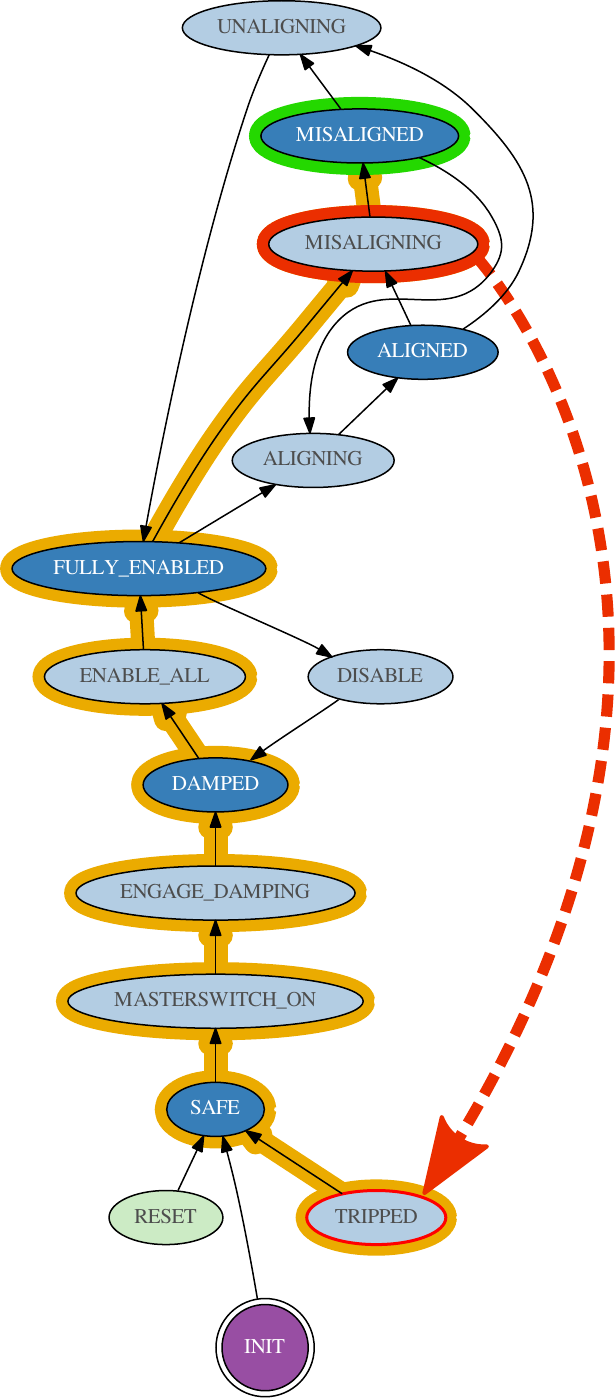}
    \caption{jump transition}
    \label{fig:graph:jump}
  \end{subfigure}
  \caption{State graph of the Advanced LIGO suspension (SUS) Guardian
    systems.  The colored ovals represent states of the system, and
    the arrows connecting states represent allowable transitions
    between states.  (a) Directives take the form of state
    \textit{requests} (green halo).  Guardian calculates the shortest
    path through the graph (orange halos) to reach the requested
    state.  (b) Any state may return a \textit{jump target}, which
    causes Guardian to immediately transition to that state.}
  \label{fig:graph}
\end{figure}

The user code for each node is a Python module that defines state
classes describing the action for each state (see
section~\ref{sec:code}) and a list of tuples that represent directed
edges---or allowable transitions---between states.  When the module is
loaded the states and edges are extracted and assembled into a
\textit{state graph} for the given system.  An example of a state
graph for the Advanced LIGO suspension system is shown in
Figure~\ref{fig:graph}.

Directives to individual nodes take the form of a \textit{state
  request}.  When a state request is received by a \texttt{guardian}
process (via its EPICS control interface) the process calculates the
shortest path in the state graph between the current state and the
requested state using a standard Dijkstra
algorithm~\cite{Dijkstra_1959}.  An example path is shown in
Figure~\ref{fig:graph:path}.  Once the currently executing state
indicates that it is complete, the process immediately transitions to
the next state in the path and begins executing the new state's code.
This repeats until the requested state is reached.

At any time, the currently executing state code may return the name of
a particular state, indicating that the system should immediately
transition to the returned state.  This is known as a \textit{jump
  transition} and is used to bypass the normal edge dynamics of the
graph.  Jump transitions allow Guardian to respond immediately to
undesired or unexpected changes in the plant.  A jump transition is
illustrated in Figure~\ref{fig:graph:jump}.

Guardian has three operating modes that determine how the system graph
is traversed:
\begin{description}[leftmargin=\parindent] \item[auto mode] Graph traversal follows the shortest path to the
  requested state.  After jump transitions the system attempts to
  automatically recover back to the requested state by following the
  path from the jump target to the previous requested state.
\item[managed mode] Graph traversal follows the shortest path to the
  requested state, as in auto mode.  After jump transitions, however,
  the system ``stalls'' at the jump target and does not transition
  away from the jump target until a new request is issued.  This mode
  is used when the node is being managed by another node, since it
  gives the managing node the opportunity to identify that the
  subordinate has jumped and redirect the subordinate to a different
  request state if needed.
\item[manual mode] The graph as a whole is ignored and the system
  immediately transitions between requested states.  This mode is
  useful only for commissioning and debugging.
\end{description}
The default mode of operation is auto mode, and nodes are
automatically put into managed mode when they are assigned a managing
node.

\subsection{State code programming and execution}
\label{sec:code}

\begin{figure}
  \begin{subfigure}[b]{\columnwidth}
    \includegraphics[width=.8\columnwidth]{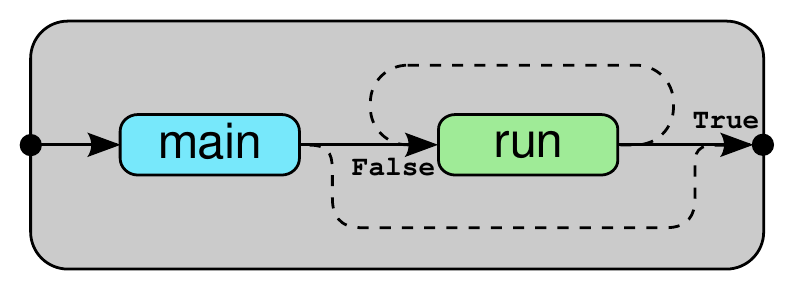}
    \caption{Basic operation, where a return value of \texttt{True}
      from either method causes the state to exit, and a return value
      of \texttt{False} causes the \texttt{run()} method to be
      executed again.}
  \end{subfigure}
  \begin{subfigure}[b]{\columnwidth}
    \includegraphics[width=.8\textwidth]{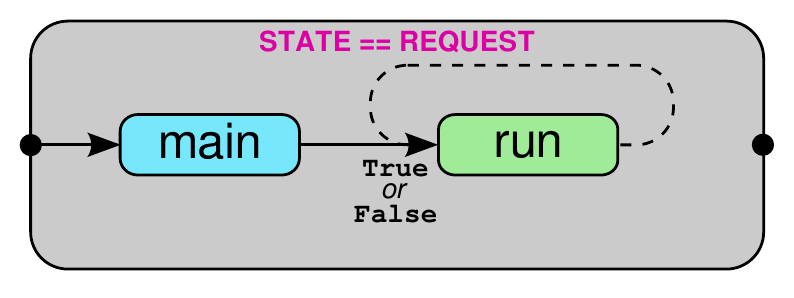}
    \caption{If the current state equals the requested state the
      \texttt{run()} method is executed again regardless of whether
      the method returns \texttt{True} or \texttt{False}.}
  \end{subfigure}
  \begin{subfigure}[b]{\columnwidth}
    \includegraphics[width=.8\columnwidth]{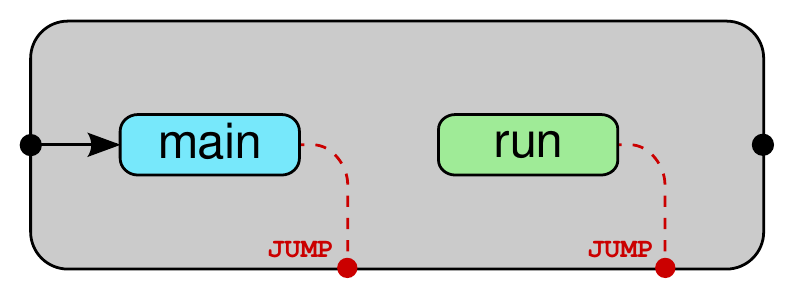}
    \caption{If either method returns a \texttt{str} state name a jump
      transition to that state occurs immediately.}
  \end{subfigure}
  \caption{Guardian state process flow, showing execution logic of the
    two state methods: \texttt{main()} and \texttt{run()}.}
  \label{fig:flow}
\end{figure}

Each Guardian state is a Python class definition that inherits from a
\texttt{GuardState} base class.  The \texttt{GuardState} class has two
methods: \main\ and \run.  The \main\ method is executed once upon
entering the state, after which the \run\ method is executed in a
loop.  The \main\ method is typically used for executing primary plant
changes, while the \run\ method is typically used to watch for state
exit conditions.  Both state methods have three return type options:
\begin{itemize}
\item \texttt{False} or \texttt{None}: This indicates that the state
  is not complete and that the current state's \run\ method should be
  executed again.
\item \texttt{True}: This indicates that the state is complete.  If
  the current state is not the target state, e.g. there are more
  states in the path, a transition is made to the next state in the
  path.  If the current state \textit{is} the target state the current
  state's \run\ method is executed again.
\item \texttt{str}: If the return value is a string it is assumed to
  be the name of a state in the graph.  The named state is set to be
  the new target state and an immediate transition to that state is
  initiated (\textit{jump transition}).
\end{itemize}
Figure~\ref{fig:flow} shows diagrams for the state method execution
logic under various conditions.

\begin{figure}
\begin{lstlisting}[language=Python]
from guardian import GuardState

class SAFE(GuardState):
    def main(self):
        ezca['SP0'] = 0
        ezca['SP1'] = 0
        self.t = ezca['T']

class DAMPED(GuardState):
    def main(self):
        ezca['SP0'] = 2
        ezca['SP1'] = 3

    def run(self):
        if ezca['RB0'] <= self.t:
            return True

edges = [
    ('SAFE', 'DAMPED'),
]
\end{lstlisting}
    \begin{picture}(0,0)
      \put(35,90){\includegraphics[width=.34\columnwidth]{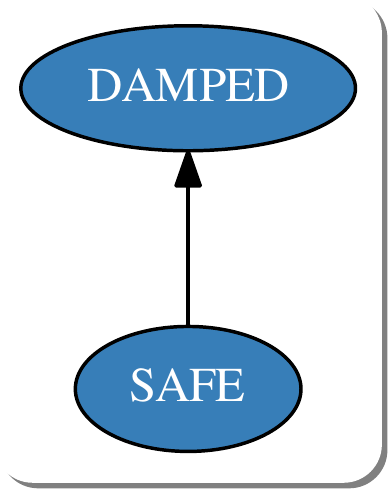}}
    \end{picture}
    \caption{Example Guardian user code module defining two states,
      SAFE and DAMPED, and one edge connecting the SAFE state to the
      DAMPED state.  The global \texttt{ezca} object is the specially
      designed LIGO EPICS client interface.  Class attributes can be
      used to pass variables between method calls.  The resulting
      system graph for this module is shown in the inset.}
    \label{fig:example_code}
\end{figure}

Figure~\ref{fig:example_code} shows an example Guardian user code
module and the resulting state graph.  The primary activity in each
state consists of monitoring the plant via \EPICS\ readback channels
and controlling the plant by writing to plant settings channels.  A
special \EPICS\ client interface object (\texttt{ezca}) provides
methods specifically designed to deal with the Advanced LIGO front end
system.

\subsection{Process architecture}

Guardian utilizes a soft real-time model for user code execution that
is similar in some respects to programmable logic controllers (PLCs).
The execution/scan loop is nominally 16~Hz, but each state method is
allowed to take as long as it needs to complete.  Once a method
returns, the next method execution occurs on the next ${1}/{16}$
second boundary.  Only one method is executed per cycle.

\begin{figure}
  \includegraphics[width=0.9\columnwidth]{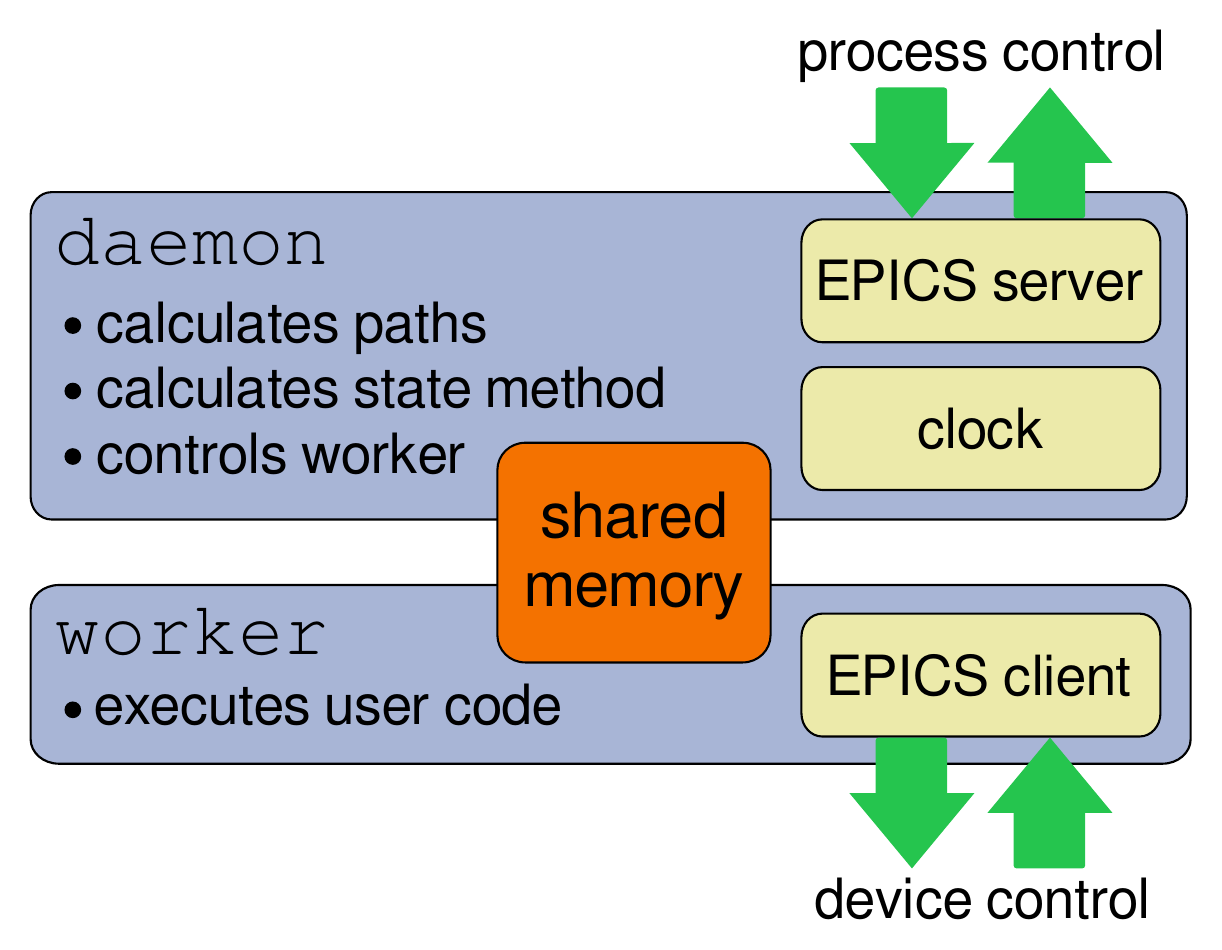}
  \caption{Architecture of the core \texttt{guardian} program.  The
    main daemon process loads the system graph, calculates
    state/method logic, and handles user interaction via a built-in
    \EPICS\ control server.  The daemon spawns the worker subprocess
    to execute all user code state methods.  All communication between
    daemon and worker is handled via a shared memory interface.}
  \label{fig:proc}
\end{figure}

In order to handle user code execution in a fully controlled
environment, the \texttt{guardian} program is designed around a
daemon/worker subprocess architecture (see Figure~\ref{fig:proc}).
The main daemon process handles operator interaction via a built-in
\EPICS\ server control interface, and determination of the appropriate
state and state method to be executed.  The worker subprocess handles
all user code execution.  The daemon and worker process communicate
via a shared memory interface.  This subprocess architecture allows
the main daemon process to terminate user code execution at any time
by simply terminating the worker subprocess.

The daemon's \EPICS\ server provides a standard interface through
which operators or other managing guardian nodes can request states in
the graph, request reload of the user code, and monitor state
execution.  When instructed from the operator, the daemon loads the
user code, constructs the state graph, and launches the worker
subprocess if it's not already running.  Through the shared memory
interface the daemon tells the worker process which state object to
instantiate and which state method to execute.  The worker then
executes the method, waits for it to return, and reports the return
value to the daemon.  The daemon then calculates the next state/method
to be executed and the process continues.  User code errors and
exceptions are captured by the worker process and reported to the
daemon, which halts further execution until the error condition is
acknowledged and cleared by the operator.

The \texttt{ezca} \EPICS\ client interface used in the worker process
keeps track of all active \EPICS\ channel subscriptions.  Channel
connectivity issues are reported to the daemon and user code can be
suspended until all connections errors have cleared.  The values of
all \EPICS\ channel writes are recorded in the \texttt{ezca} object
and a set point monitor can be used to check settings during each
execution cycle.  This allows for tracking of set points and
notifications if any set point has changed out of band.

\subsection{Inter-node management}

A special \texttt{NodeManager} interface is provided to facilitate
inter-node control.  Manager nodes list their subordinate nodes in the
\texttt{NodeManager} object.  Once the subordinate nodes are
``acquired'' the \texttt{NodeManager} object sets the subordinate
nodes to be in managed mode and starts tracking their state and status
channels.  Via the interface, manager nodes can request states of
their subordinates and inspect their state and status in an idiomatic
way.

\begin{figure*}[ht!]
  \begin{subfigure}[b]{0.74\textwidth}
    \includegraphics[width=\textwidth]{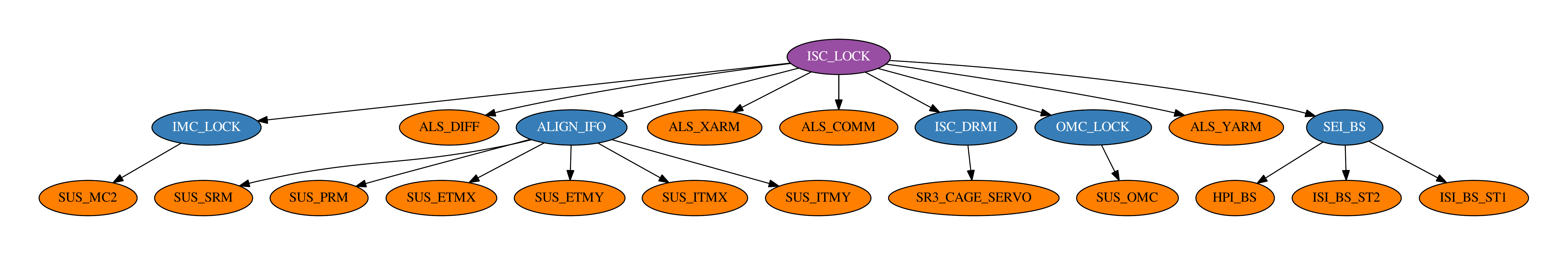}
    \caption{\node{ISC\_LOCK} node hierarchy}
    \label{fig:hier:LOCK}
  \end{subfigure}
  \begin{subfigure}[b]{0.25\textwidth}
    \includegraphics[width=\textwidth]{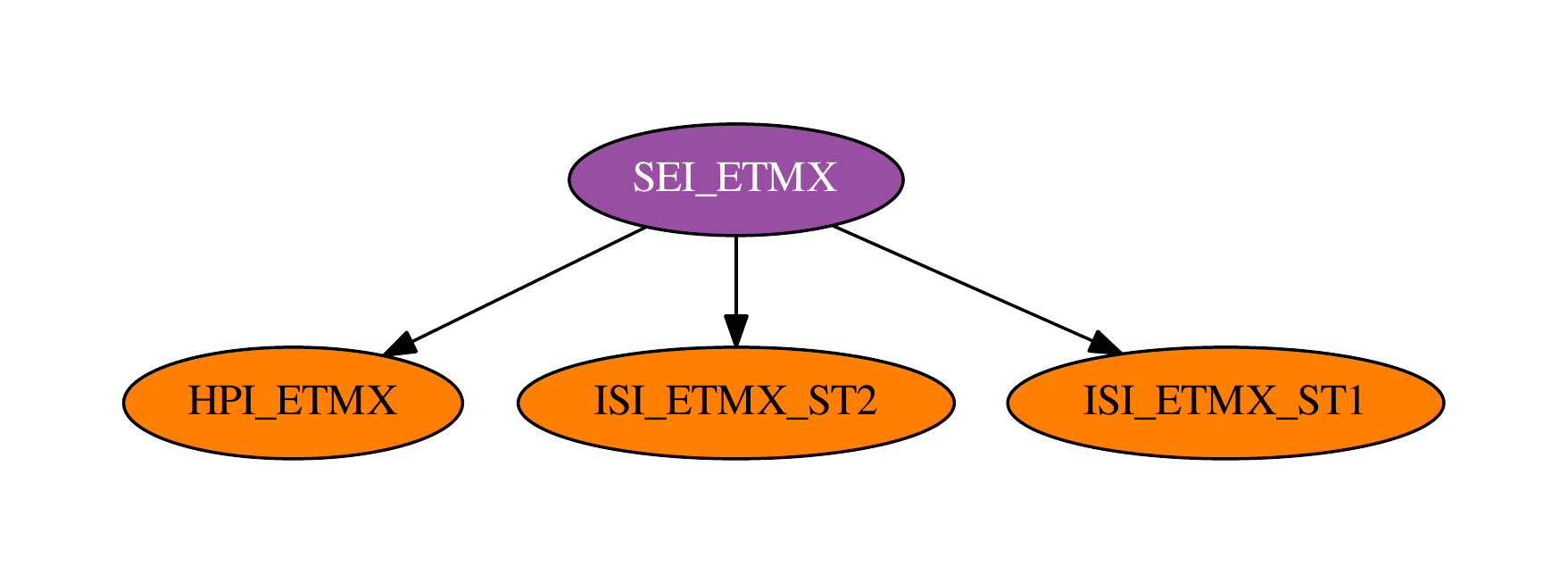}
    \caption{SEI nodes}
    \label{fig:hier:SEI}
  \end{subfigure}
  \caption{Guardian node hierarchy for various Advanced LIGO
    subsystems.  (a) node hierarchy below the \node{{ISC\_LOCK}} node
    that handles the interferometer lock acquisition process.  (b) An
    SEI node hierarchy (for \node{SEI\_ETMX} specifically) for which
    there are 11 similar in the full interferometer control.  The top
    node for this system is known as the ``chamber manager''.}
  \label{fig:hier}
\end{figure*}

\subsection{State tracking and validation}

Instrument state tracking is critical for scientific applications
where instrument validation needs to be well documented during an
experiment, and there are a couple of key benefits of the state
machine approach that make it particularly well suited for this task.
First, at any given time the system can be in only a single
well-defined state, which helps eliminate ambiguities.  Second, since
there are no persistent variables between states, all process
variables of the system must be external to the state machine itself.
In the case of Guardian this is achieved by having all process
variables be \EPICS\ records that are archived by the data acquisition
system.  Since Guardian node states and status are also broadcast over
\EPICS\ and recorded by the data acquisition system, the complete
state of the instrument at any given point in time can be
reconstructed completely from data on disk.

Guardian has additional features to facilitate instrument validation
during operation.  A \textit{nominal} state can be defined for each
Guardian node indicating the state the system is expected to be in
during nominal operation.  Each Guardian node can then broadcast the
overall status of the system via a single binary status channel.  The
conditions that are checked are:
\begin{enumerate*}[label={\alph*)}]
\item the requested state is equal to the nominal state,
\item the current state is equal to the nominal state,
\item the node is in execution mode (not paused or stopped), and
\item there are no error conditions present.
\end{enumerate*}
If all of those conditions are met, the node returns an overall status
of ``OK'' to indicate that the system is ready for operation.

\subsection{User interface features}

All user code can be reloaded on the fly on a live system, even while
in the middle of state execution.  A snapshot of all user code at time
of load is committed into per-node user code git archives, which
allows for inspecting the exact code that was running on any node at
any point in time.

A notification system provides a way for state code to push important
notifications to the operator.  Verbose logs are archived by a logging
infrastructure that provides full access to all node logs over time.

A supporting suite of utilities is available to draw state graphs,
analyze and validate code, etc.  Further details, installation and
usage instructions, and a description of code syntax can be found
in the LIGO document control center~\cite{dcc:T1500292}.

\section{Guardian supervision of Advanced LIGO}

Advanced LIGO employs roughly 100 Guardian nodes in the full control
of each detector (a number that continues to increase as new
subsystems and functionality are commissioned).  Table~\ref{tab:nodes}
shows the breakdown of nodes among the various subsystem components.
Figure~\ref{fig:hier} shows graphs of the node hierarchies for the ISC
and SEI BSC subsystems.

\begin{table}
  \begin{tabular*}{\columnwidth}{@{\extracolsep{\fill}}l|c|c|c}     subsystem
    & number of units & nodes per unit & total nodes \\
    \hline
    ISC
    & 1               & 9              &  9 \\
    SUS
    & 26              & 1              & 26 \\
    SEI
    & 10              & 4/3            & 35 \\
        TCS
    & 2               & 2              &  4 \\
    DIAG
    & 4               & 1              &  4
  \end{tabular*}
  \caption{Advanced LIGO Guardian node breakdown among subsystems.
    ISC: interferometer sensing and control; SUS: suspensions; SEI:
    seismic isolation; TCS: thermal compensation; DIAG: diagnostics.}
  \label{tab:nodes}
\end{table}

\subsection{Subsystem control}

The Advanced LIGO Guardian implementation takes advantage of
standardization in subsystem hardware and the accompanying real-time
controls code to create a highly modular and distributed automation
infrastructure.

The suspension subsystem defines a single Guardian module to describe
automation for all suspension systems in the interferometer.  A common
suspension class object abstracts various suspension readout and
control functions depending on the suspension type.  The Guardian code
for each individual suspension system merely specifies the suspension
type, and an appropriate \EPICS\ channel access prefix, then loads the
common suspension state graph.  All suspension systems in Guardian
therefore present an identical state graph interface to the rest of
the system.

The seismic isolation subsystem further modularizes its code among the
different types of control loops employed by the various ISI and HEPI
devices.  Sub-packages define functions, states and sub-graphs
separately for damping and isolation control, as well as for system
initialization and watchdog handling.  The full system graph for each
SEI component is then assembled from the necessary components.
Additionally, due to the complexity of interaction between the various
isolation stacks on a given chamber, multiple Guardian nodes are used
to cover the SEI systems for a single chamber.  For the larger BSC
chambers that house the beam splitter and core test mass optics, one
node handles the HEPI system, two nodes handle the two stages of the
ISI system, and a ``chamber manager'' is used to orchestrate their
actions.  This hierarchy can be seen in Figure~\ref{fig:hier:SEI}.
The smaller HAM chambers that house auxiliary optics use a similar
hierarchy except with only a single ISI node for the single stage ISI
system.

The SUS and SEI Guardian systems constantly monitor the state of their
plant watchdogs.  If a watchdog trip occurs, the systems will
immediately jump to special states that reset all control loops and
wait for the operator to reset the watchdogs.  Once the watchdogs are
reset, Guardian automatically brings all systems back to their
previously requested states.

Overall interferometer lock acquisition is handled by the hierarchy of
nodes shown in Figure~\ref{fig:hier:LOCK}.  Separate nodes handle
locking the IMC, OMC, and ALS systems, and the overall DC alignment of
the various suspension controllers.  The full lock acquisition
process~\cite{Staley_2014} is orchestrated by the \node{ISC\_LOCK}
node.

\subsection{System diagnostics and system validation}

A set of specialized diagnostic (DIAG) nodes are employed to monitor
aspects of the instrument that are not directly handled by the primary
automation and subsystem nodes.  These look at things like laser
status, light levels on various detectors, the states of various
electronics modules, the state of the Beckhoff system, etc.

At the top of the instrument node hierarchy is an ``IFO top node''
whose job is to monitor the status of all other nodes in the system.
This node provides a single channel that reports on the status of the
entire system as a whole, which is critical in determining if the
observatory is ready to begin observation or not.

The status reporting of the Guardian system is also used extensively
for detector characterization and validation purposes.  Downstream
detector characterization processes use individual subsystem status
reporting during analysis of subsystem behavior.

\subsection{Performance}

The first operational demonstration of the full Guardian deployment in
LIGO was the first Advanced LIGO observing run from September 2015 to
January 2016, during which LIGO made the first ever direct detection
of gravitational waves.  The system performed robustly with no issues
of note, and was used to aid in validation of the first gravitational
wave event candidates.

\begin{figure}
  \includegraphics[width=\columnwidth]{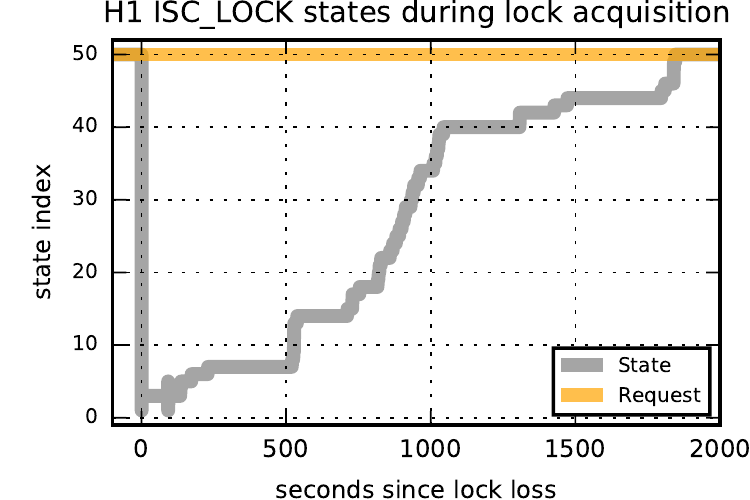}
  \caption{An example of the \node{ISC\_LOCK} Guardian node at the H1
    detector autonomously recovering the interferometer to full
    low-noise lock after an unintended lock loss.  The y-axis shows
    arbitrary state indices for the node.  Before $t=0$ the system is
    in the nominal locked state, corresponding to the orange
    ``Request'' state.  At $t=0$ the system loses lock and the system
    transitions to a ``DOWN'' state that resets all control loops in
    preparation for re-acquisition.  At the end, the system has
    recovered the initial nominal lock state.  In this particular
    instance, there is at no point any human intervention.  The
    recovery time in this case is roughly 30 minutes.}
  \label{fig:lock}
\end{figure}

The fully commissioned lock acquisition process takes about 30
minutes, with some variability between the two LIGO detectors.
Figure~\ref{fig:lock} shows the primary lock acquisition node
(\node{ISC\_LOCK}, top node in Figure~\ref{fig:hier:LOCK}) at the
Hanford ``H1'' detector as it progresses through the states in the
lock acquisition sequence during a fully autonomous lock loss
recovery.  The limiting factors in the recovery time are generally the
physical responses of the various interferometer subsystems during the
engagement of various control loops and not by anything inherent to
the Guardian system itself.  Lock acquisition time will likely be
improved with further commissioning.

\begin{figure}
  \includegraphics[width=\columnwidth]{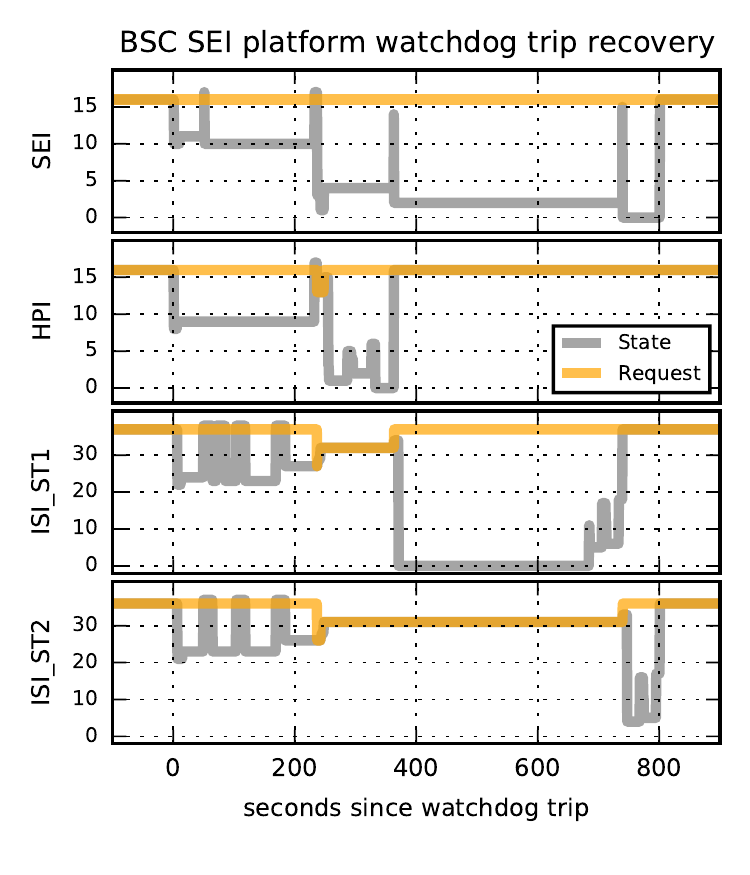}
  \caption{An example of watchdog recovery for a BSC seismic system
    (see Figure~\ref{fig:ifo}).  The y-axis shows arbitrary state
    indices for each node.  The top row is the \node{SEI} ``chamber
    manager'' node which manages the actions of the following nodes
    shown in the lower three plots: \node{HPI}, for the BSC HEPI
    system; \node{ISI\_ST1}, for the first stage of the BSC ISI
    system; \node{ISI\_ST2}, for the second stage of the BSC ISI
    system.  The recovery procedure is described in
    Section~\ref{sec:sei}.  The full recovery time is roughly 13
    minutes.}
  \label{fig:isolation}
\end{figure}

Figure~\ref{fig:isolation} shows the progression of states for the
hierarchy of nodes controlling the seismic isolation system in a BSC
chamber during autonomous recovery from a watchdog trip.  During
normal operation, these types of trips are usually the result of
earthquakes.  The nodes involved are the same as those shown in the
hierarchy in Figure~\ref{fig:hier:SEI}.  The chamber manager node
coordinates the activity of three subordinate nodes by issuing state
requests.  The watchdog recovery time is less than 15 minutes.  The
distributed nature of Guardian allows all suspension and seismic
systems to recover in parallel if multiple trips occur simultaneously,
thereby significantly reducing overall recovery time.

\section{Conclusion and outlook}

Guardian has proven to be a powerful tool for commissioning the
Advanced LIGO detectors and has demonstrated robustness in the
operation of both detectors during the first Advanced LIGO observing
run.  Many future runs are planned and continual improvement will be
made to the user code logic during commissioning breaks.
Additionally, short term detector improvements will involve new
automation challenges.  In particular, plans are being made to
increase laser power in the interferometers, which will increase the
complexity of the lock acquisition process and require full
commissioning of the thermal compensation system, as well as
potentially introducing additional nodes to handle issues such as
parametric instabilities~\cite{Evans_2015}.  Longer term plans involve
even more substantial upgrades that will introduce new subsystems with
their own automation requirements, such as squeezed light systems to
reduce quantum noise~\cite{Miller_2015}.

While the Guardian platform itself is now stable, many new features
are planned for future releases.  Further integration of the user code
archive will facilitate detailed inspection of the historical data for
detector characterization purposes.  User interface improvements are
planned that will integrate system graphs directly into the control
UI.  Node management interfaces will be streamlined.

Success in LIGO is also leading to the adoption of Guardian by other
long-baseline gravitational wave detectors.  Guardian is being used by
the Japanese 3~km-baseline, cryogenic gravitational-wave detector
\mbox{KAGRA}~\cite{Aso_2013}, and may be used by Advanced Virgo.
Plans are also afoot to install a third LIGO detector in India.

\section*{Acknowledgments}

The author would like to thank the following people: Daniel Sigg for
many fruitful discussions on the theory and practice of automation in
general and automation of gravitational wave detectors in particular,
Matthew Evans and Sam Waldman for their initial seed of an idea and
subsequent sprout of work, Charles Celerier for his invaluable help in
early development and testing, and Robert Ward for breaking the ice
and getting the commissioners using this new system.  In addition, the
author thanks the entire Advanced LIGO commissioning team who put up
with the initial growing pains, helped push the system to its full
potential, and wrote most of the user code that actually controls
these incredible instruments.

LIGO was constructed by the California Institute of Technology and
Massachusetts Institute of Technology with funding from the National
Science Foundation and operates under Grant No. PHY-0757058. Advanced
LIGO was built under award PHY-0823459.

\newpage

\bibliography{paper}

\end{document}